\documentstyle{article}
\topmargin = - 0.5 cm
\begin{document}
\setlength{\baselineskip}{1\baselineskip}
\title{Simple New Axioms for Quantum Mechanics}
\author{N.P. Landsman\thanks{E.P.S.R.C. Advanced Research Fellow.
 E-mail: NPL@AMTP.CAM.AC.UK.
} \\ \mbox{} \hfill \\ Department of Applied Mathematics and
Theoretical Physics\\ University of Cambridge\\ Silver Street,
Cambridge CB3 9EW, U.K.}
\date{\today}
\maketitle
\begin{abstract}
The space ${\cal P}$ of pure states of any physical system, classical
or quantum, is identified as a Poisson space with a transition
probability. The latter is a function $p:{\cal P}\times{\cal
P}\rightarrow [0,1]$; in addition, a Poisson bracket is defined for
functions on ${\cal P}$. These two structures are connected through
unitarity. Classical and quantum mechanics are each characterized by a
simple axiom on the transition probability $p$.  Unitarity then
determines the Poisson bracket of quantum mechanics up to a
multiplicative constant (identified with Planck's constant).
Superselection rules are naturally incorporated.
\end{abstract}
\section{Introduction}
Axiomatic quantum mechanics (cf.\ \cite{Gud,BC} for representative
overviews) is usually inspired by a mixture of two extreme attitudes.
One the one hand, one could try to show that the Laws of Thought
necessarily imply that Nature has to be described by quantum
mechanics. On the other hand, quantum mechanics could be a contingent
theory. In this Letter I will show that quantum mechanics can be
described by one axiom that is fairly general, incorporates classical
mechanics as well, and may fall into the first category, and by two
further axioms which, in my opinion, are clearly contingent.

 The purpose of my axiomatization is twofold. Firstly, it suggests at
what point quantum mechanics may be modified. Secondly, it formalizes
classical and quantum mechanics in parallel, so that it becomes
crystal-clear to what extent these two theories agree, and where they
(dramatically) differ. Thus I expect the structure set out below to be
useful in the theory of quantization as well as of the classical limit
of quantum mechanics.

The usual formulation of quantum mechanics in terms of linear
operators on a Hilbert space obviously does not serve this second
purpose well. However, even more refined schemes involving convex
structures and/or quantum logic \cite{Gud,BC} are of limited use in
the present context: apart from their (arguable) lack of intuitive
appeal, such approaches miss the Poisson structure that plays such a
crucial role in mechanics \cite{MR}. Nonetheless, the mathematical
proofs that are necessary to show that the axioms proposed below
indeed describe quantum mechanics use these schemes in an essential
way.
\section{Poisson brackets}
The pure states of a classical mechanical system are the points of its
phase space ${\cal P}$. This space is equipped with a Poisson
structure, that is, for any two (smooth) functions $f,g$ on ${\cal P}$
the Poisson bracket $\{f,g\}$ is defined. One calls ${\cal P}$ a {\em
Poisson manifold}. Thus any (smooth) function $H$ on ${\cal P}$
defines a {\em Hamiltonian vector field} $X_H$ on ${\cal P}$ by
$X_H(g)=\{H,g\}$, and the Hamiltonian equations of motion satisified
by a curve $\sigma(t)$ in ${\cal P}$ are \cite{MR}
\begin{equation}
\frac{d\sigma(t)}{dt}=X_H(\sigma(t)). \label{ham}
\end{equation}

 In the sixties it was discovered by many people (cf.\ \cite{MR} for a
modern presentation and references) that quantum mechanics may, to
some extent, be brought into the same form.  Here one chooses ${\cal
P}={\sf P}({\cal H})$, the projective space of ${\cal H}$, the Hilbert
space of (pure) states of the system. Every Hermitian linear operator
$\hat{A}$ on ${\cal H}$ defines a real-valued function $f_A$ on ${\cal
P}$ by \begin{equation}
f_A(\psi)=\langle\psi|\hat{A}|\psi\rangle/\langle\psi|\psi\rangle,
\end{equation}
where $\psi\in{\sf P}({\cal H})$ is the image of $|\psi\rangle\in{\cal
H}$. The Poisson bracket of such functions is essentially given by the
commutator:
\begin{equation}
\{f_A,f_B\}=\frac{i}{\hbar}f_{[\hat{A},\hat{B}]}. \label{pb}
\end{equation}
The Schr\"{o}dinger equation (projected to ${\sf P}({\cal H})$) is
then precisely (\ref{ham}), with $H=f_H$.

Hence quantum mechanics may be described in the language of classical
mechanics, with some curious extra rules: the phase space is ${\cal
P}={\sf P}({\cal H})$, the Poisson bracket is defined by (\ref{pb}),
and only functions of the form $f_A$ (rather than all smooth functions
on ${\cal P}$, as in classical mechanics) correspond to observables
(but note that the Poisson bracket of any two functions on ${\cal P}$
is determined by the special case (\ref{pb})).

In order to formulate the axioms below, I need to recall a basic
theorem on Poisson structures
\cite{MR}. Namely, every Poisson manifold ${\cal P}$ can be decomposed as the union of its {\em symplectic
leaves}: these are maximal subspaces on which the Poisson structure is
non-degenerate. This means that at each point $\rho$ the Hamiltonian
vector fields $X_f$ span the tangent space of the leaf through $\rho$.
\section{Transition probabilities}
It was known from the earliest days of quantum mechanics that the
notion of a transition probability is of central importance to this
theory. Abstractly, a transition probability $p$ on a set ${\cal P}$
is a function on ${\cal P}\times {\cal P}$, taking values in the
interval $[0,1]$, with the special property that $p(\rho,\sigma)=1$ is
equivalent to $\rho=\sigma$ \cite{vN,BC}. Moreover, in general one
assumes the symmetry property $p(\rho,\sigma)=p(\sigma,\rho)$. In
standard quantum mechanics one puts ${\cal P}={\sf P}({\cal H})$, as
above, and \begin{equation} p(\rho,\sigma)=|\langle
R|\Sigma\rangle|^2, \label{qtp}
\end{equation}
where $|R\rangle$ and $|\Sigma\rangle$ are unit vectors in ${\cal H}$
which project to $\rho$ and $\sigma$ in ${\sf P}({\cal H})$,
respectively. This choice of $p$ is essentially the Born rule. Note
that $p(\rho,\sigma)=f_{[\sigma]}(\rho)$, where the linear operator
$\widehat{[\sigma]}$ on ${\cal H}$ is the orthogonal projector onto
the one-dimensional subspace spanned by $|\Sigma\rangle$.

The physical meaning of transition probabilities implies that in the
case of classical mechanics, where ${\cal P}$ is an arbitrary
manifold, one has to put
\begin{equation}
p(\rho,\sigma)=\delta_{\rho\sigma}\:\forall\rho,\sigma. \label{ctp}
\end{equation}

In what follows I need the (obvious) result that any space with a
transition probability decomposes as the union of its irreducible
components, called {\em sectors} (a subspace is irreducible if it is
not the union of two mutually orthogonal spaces). Also, there are some
technical requirements on transition probabilities (each maximal
orthogonal subset being a basis, and orthoclosed subsets being
subspaces, cf.\ \cite{BC} and \cite{NPL}) that I shall assume.

The superposition principle of quantum mechanics (which is normally
expressed in terms of vectors in a Hilbert space) can be described in
the present language. For any subset $Q$ of ${\cal P}$ one defines the
orthoplement
\begin{equation}
Q^{\perp}=\{\sigma\in{\cal P}|p(\rho,\sigma)=0\: \forall \rho\in Q\}.
\label{ortho}
\end{equation}
The possible superpositions of the pure states $\rho,\sigma$ are then
the elements of $\{\rho,\sigma\}^{\perp\perp}$. If $\rho$ and $\sigma$
lie in different sectors then clearly
$\{\rho,\sigma\}^{\perp\perp}=\{\rho,\sigma\}$.  \section{The axioms}
The mathematical structure characterizing pure state spaces in
classical and quantum mechanics can now be identified.  A {\em Poisson
space with a transition probability} is at the same time a transition
probability space $({\cal P},p)$ and a Poisson manifold $({\cal P},
\{\, ,\,\})$, such that the Poisson structure is {\em unitary} in the
following sense.  For each $\rho\in{\cal P}$ I define a function
$p_{\rho}$ on ${\cal P}$ by $p_{\rho}(\sigma)=p(\rho,\sigma)$. Using
$p_{\rho}$ as the Hamiltonian $H$, the Hamiltonian flow $\sigma(t)$ on
${\cal P}$ is given by the solution of (\ref{ham}).  Unitarity now
means that for each $\rho$ this flow leaves the transition
probabilities invariant, in that
$p(\sigma_1(t),\sigma_2(t))=p(\sigma_1,\sigma_2)$ for all
$\sigma_1,\sigma_2\in{\cal P}$ and all $t$.
 
My axioms on the pure state space ${\cal P}$ of quantum mechanics with
(discrete) superselection rules are:
\begin{itemize}
\item
QM1: The pure state space ${\cal P}$ is a Poisson space with a
transition probability;
\item
 QM2: For each pair $(\rho,\sigma)$ of points which lie in the same
sector of ${\cal P}$, $\{\rho,\sigma\}^{\perp\perp}$ is isomorphic to
${\sf P}({\bf C}^2) $ as a transition probability space; \item QM3:
The sectors of $({\cal P},p)$ coincide with the symplectic leaves of
$({\cal P}, \{\, ,\,\})$.
\end{itemize}
Here ${\sf P}({\bf C}^2)$ is understood to be equipped with the usual
Hilbert space transition probabilities.  Axiom QM2 was inspired by a
mathematical paper on convexity theory and operator algebras
\cite{AHS}.

To characterize classical mechanics, one simply postulates the axioms
CM1 = QM1, and CM2 = eq.\ (\ref{ctp}). In this case,
$\{\rho,\sigma\}^{\perp\perp}$ simply equals $\{\rho,\sigma\}$, and
each point is a sector.
\section{Consequences of the axioms}
 As outlined in the next section, it can be shown  
that the axioms QM1-QM3 imply that ${\cal P}=\cup_i {\sf P}({\cal
H})_i$ (which is meant as a union over sectors).  Here each ${\cal
H}_i$ is a Hilbert space, and the transition probabilities in each
sector ${\sf P}({\cal H})_i$ are given by (\ref{qtp}).  Moreover, the
Poisson bracket on ${\cal P}$ is determined up to a multiplicative
constant $\hbar$, and is such that in each sector (or, equivalently,
symplectic leaf) ${\sf P}({\cal H})_i$ it is given by (\ref{pb}). I
find it satisfying to see Planck's constant enter as a free parameter
allowed by the axioms.

This illustrates a remarkable difference between classical and quantum
mechanics: in the former the pure state space and the Poisson
structure can be freely specified, whereas in the latter the only
freedom lies in the dimensions of the ${\cal H}_i$ and the (nonzero)
value of $\hbar$.

I now show how the usual observables of quantum mechanics can be
reconstructed, restricting myself to the finite-dimensional case (see
\cite{NPL} for the general construction, which also applies to
classical mechanics). Firstly, the space of observables is simply the
real vector space ${\cal A}({\cal P})$ of (real) finite linear
combinations of the functions $p_{\rho}$, where $\rho$ runs through
${\cal P}$. In other words, the observables of quantum mechanics are
in essence the transition probabilities.  Secondly, one has a spectral
theorem in ${\cal A}({\cal P})$: every function $f=\sum \mu_i
p_{\rho_i}$ can be rewritten as $f=\sum_j \lambda_j p_{e_j}$, where
$p(e_j,e_k)=\delta_{jk}$. This given us a squaring map $f^2:=\sum_j
\lambda^2_j p_{e_j}$, and subsequently a commutative product (Jordan
structure) by
\begin{equation}
f\circ g= {\mbox{\footnotesize $\frac{1}{4}$}} ((f+g)^2-(f-g)^2),
\end{equation}
which is bilinear because of the special form (\ref{qtp}) of $p$.  I
now complexify ${\cal A}({\cal P})$, and define a product $\cdot$ on
${\cal A}({\cal P})_{\bf C}$ by
\begin{equation}
f\cdot g=f\circ g-{\mbox{\footnotesize $\frac{1}{2}$}}i\hbar \{f,g\}.
\end{equation}
This product turns out to be associative as a consequence of the
unitarity relating the transition probability (which is ultimately
responsible for the product $\circ$) and the Poisson bracket.
Finally, one (easily) shows that the algebra $({\cal A}({\cal P})_{\bf
C},\cdot)$ thus constructed is a direct sum of matrix algebras.
Indeed, in case of a single sector the usual spectral theorem already
says that any function $f_A$ lies in ${\cal A}({\cal P})$ (for
Hermitian $\hat{A}$). In any case, it is pleasant to represent
observables as real-valued functions on the space of pure states, just
like in classical mechanics.  \section{Outline of the proof} I here
give the main steps in showing that the axioms imply that ${\cal
P}=\cup_i {\sf P}({\cal H})_i$, with Poisson structure as indicated.
For full details see \cite{NPL}; lattice-theoretic definitions may be
found in \cite{Kal} or \cite{BC}.  For simplicity I restrict myself
here to the irreducible case of one sector.

Firstly, on the basis of Axiom QM1 one constructs a complete atomic
orthomodular lattice ${\cal L}({\cal P})$ \cite{BC}, whose members are
the orthoclosed subspaces of $\cal P$. Axiom QM2 then leads to the
conclusion that this lattice has the covering property (i.e.,
satisfies the exchange axiom); the proof is by induction on the
dimension of the lattice. One then uses the (von Neumann)
coordinatization procedure for projective lattices \cite{BC}, which
leaves one with an arbitrary division ring ${\bf D}$. {\em This step
can only be performed if the dimension of $\cal P$ (as a transition
probability space) is not equal to 3, and this limitation restricts my
result.} One then uses Axiom QM2 once again to prove that ${\bf
D}={\bf C}$; surprisingly, this is the most difficult step. Standard
theorems \cite{BC} then lead to the conclusion that the transition
probability must come from the usual inner product on a complex
Hilbert space $\cal H$. Hence ${\cal P}={\sf P}({\cal H})$.

 I now come to the identification of the Poisson structure on $\cal
P$.  Axiom QM3 implies that each sector is a symplectic space.
Unitarity (in my sense) and Wigner's theorem (cf.\ \cite{BC}) imply
that each $f_A$ generates a flow on ${\sf P}({\cal H})$ which is the
projection of a unitary flow on $\cal H$. Therefore,
$\{f_A,f_B\}(\sigma)=\frac{d}{dt}f_B(\exp(itC(A))\sigma)_{t=0}$ for
some Hermitian operator $\hat{C}$ on $\cal H$, depending on $A$ (here
$\exp(itC(A))\sigma$ is the projection of
$\exp(it\hat{C}(A))|\Sigma\rangle$, using my previous notation). The
right-hand side equals $f_{i[\hat{C},\hat{B}]}$. Anti-symmetry of the
left-hand side implies that $\hat{C}=\hbar \hat{A}$ for some $\hbar\in
{\bf R}$, and this leads to the desired result.  The multiplicative
constant $\hbar$ must be nonzero in order to satisfy Axiom QM3. In
principle, it may depend on the sector, but given that it is nonzero
this can be undone by a simple rescaling of the Poisson bracket.
\section{Beyond quantum mechanics}
In my opinion, the most remarkable aspect of these axioms lies in the
universality of the transition probabilities in quantum mechanics.
Take any quantum system, and any two of its pure states: axiom QM2
describes their superpositions and transition probabilities. This
strongly suggests that there should be some underlying explanation for
these transition probabilities. The central limit theorem of
probability theory comes to mind: whatever the individual probability
distribution, if one has a large number of replicas one will find that
fluctuations are described by the Gaussian (normal) distribution.  One
would hope that the `distribution' (\ref{qtp}) emerges in a similar
way as some universal limit.  

\begin{thebibliography}{99}
\bibitem{Gud}   S.P. Gudder,  {\em Stochastic Methods in Quantum Mechanics}
 (North-Holland, Amsterdam, 1979).
\bibitem{BC}   E.G. Beltrametti  and
                 G. Cassinelli, {\em The Logic of Quantum Mechanics}
(Cambridge University Press, Cambridge, 1984).
\bibitem{MR}  J.E. Marsden  and T.S. Ratiu, 
{\em Introduction to Mechanics and Symmetry}
(Springer, New York, 1994).
\bibitem{vN}  J. von Neumann, 
{\em Continuous geometries with a transition probability},   
Mem.\ Amer.\ Math.\ Soc. {\bf 252} (1981) (MS from 1937).
\bibitem{NPL} N.P. Landsman, quant-ph/9603005 
(submitted to Rev.\ Math.\ Phys.).
\bibitem{AHS}   E.M. Alfsen, H. Hanche-Olsen, and  F.W. Shultz,
  Acta Math. {\bf 144}, 267 (1980).
\bibitem{Kal} G. Kalmbach, {\em Orthomodular Lattices} 
(Academic Press, London, 1983).
\end{thebibliography}
 \end{document}